\documentclass[prl,aps,amssymb,twocolumn,superscriptaddress, notitlepage,longbibliography]{revtex4-2}
\usepackage{bm,amssymb,amsmath,dsfont,mathrsfs,amstext,latexsym, physics,relsize}
\usepackage[dvipsnames]{xcolor}
\newcommand{\cbar}{\overline{c}}
\definecolor{mylink}{HTML}{990000}
\usepackage[colorlinks=true,citecolor=mylink,urlcolor=mylink,linkcolor=mylink]{hyperref}
\usepackage{orcidlink}
\usepackage{float}
\usepackage{comment}   
\begin{document}

\newcommand{\dt}[2]{\operatorname{det}\nolimits_{#1}#2}

\title{Ground-State Chiral Quantum Clusters}

\author{Matteo~Massaro\orcidlink{0009-0004-0935-8022}}
\email{matteo.massaro@uni.lu}
\affiliation{Department  of  Physics  and  Materials  Science,  University  of  Luxembourg,  L-1511  Luxembourg,  Luxembourg}
\author{Adolfo~del Campo\orcidlink{0000-0003-2219-2851}} 
\email{adolfo.delcampo@uni.lu}
\affiliation{Department  of  Physics  and  Materials  Science,  University  of  Luxembourg,  L-1511  Luxembourg, Luxembourg}
\affiliation{Donostia International Physics Center,  E-20018 San Sebasti\'an, Spain}
\affiliation{Institute of Science and Engineering, Kanazawa University, Kakuma-machi, Kanazawa, Ishikawa,  920-1192, Japan}

\begin{abstract}
Quantum time crystals are states of matter that spontaneously break time translation symmetry, extending the notion of crystalline order to time. While their out-of-equilibrium version has been successfully realized by periodic driving, the equilibrium counterpart remains elusive. Here we introduce \emph{ground-state chiral quantum clusters}: self-bound many-body ground states supporting a persistent current. We provide a minimal field theory describing these states, and an exact microscopic realization via the attractive anyonic Lieb--Liniger model on a ring. In the latter, the ground-state density-density correlator exhibits the Talbot effect, with a soliton-like profile that periodically disperses and revives while drifting indefinitely along the ring, realizing persistent motion at equilibrium. Possible experimental probes are discussed.
\end{abstract}

\maketitle

The concept of quantum time crystals was introduced by Wilczek~\cite{Wilczek2012}, who proposed that a time-independent many-body Hamiltonian could possess a ground state exhibiting persistent periodic motion, thereby spontaneously breaking continuous time-translation symmetry. This suggestion extended the paradigm of spontaneous symmetry breaking into the temporal domain and generated immediate interest and debate \cite{Bruno2013Comment, Wilczek2013Reply}. 
Shortly thereafter, several no-go theorems clarified the limitations of this proposal.
Bruno demonstrated that ground-state rotation in equilibrium systems of the type envisioned in the original proposal is forbidden under broad conditions~\cite{Bruno2013}. More generally, Watanabe and Oshikawa proved that in systems with short-range interactions, neither ground states nor thermal equilibrium states can exhibit persistent time-dependent long-range order~\cite{WatanabeOshikawa2015}. These results established that continuous time-translation symmetry cannot be spontaneously broken in the thermodynamic limit in equilibrium eigenstates of generic time-independent Hamiltonians \cite{Sacha2018,Sacha2020}.

Subsequent efforts focused on identifying systems that evade the assumptions underlying the equilibrium no-go results \cite{Zaletel2023}. One option is to  rely on multiple-body interactions, potentially realizable in quantum simulators \cite{Kozin2019}. A widely explored direction has been the study of periodically driven (Floquet) systems, in which time-translation symmetry is discrete rather than continuous.
Else, Bauer, and Nayak demonstrated that interacting Floquet systems can support a phase that spontaneously breaks discrete time-translation symmetry~\cite{Else2016}. Yao \emph{et al.} provided a detailed theoretical characterization of such discrete time crystals and their stability mechanisms~\cite{Yao2017}. These theoretical advances were rapidly followed by experimental realizations. Discrete time-crystalline order was observed in trapped-ion systems~\cite{Zhang2017} and in ensembles of nitrogen-vacancy centers in diamond~\cite{Choi2017}, confirming that subharmonic temporal order can arise as a robust non-equilibrium phase. Subsequent work has shown that such behavior can be stabilized either by many-body localization or by prethermalization mechanisms. Thus, continuous time-translation symmetry cannot be broken in equilibrium ground states of generic short-range systems, while discrete time-translation symmetry can be broken in driven, excited, or metastable regimes. This distinction reshapes the conceptual landscape of temporal symmetry breaking.

In this Letter, we revisit equilibrium many-body states that exhibit persistent currents without violating the no-go constraints. Equilibrium persistent currents in quantum systems are often viewed as kinematic: flux selects a winding sector in an otherwise homogeneous fluid. In such cases, the current reflects a global phase twist rather than a qualitative restructuring of the many-body state.  A more structured situation arises when the current is carried by a self-bound quantum object, such as a droplet, whose ground state exhibits strong clustering correlations. In this regime, chirality becomes an intrinsic property of a composite quantum fluid rather than a mere boost of a uniform phase. The current is attached to a finite, correlated many-body object with internal structure and collective modes.  A persistent current can arise in a number of scenarios involving flux, spin-orbit coupling, magnetic field, or intrinsic pairing with angular momentum. However, the occurrence of a chiral persistent current in the ground state is uncommon.  Understanding such equilibrium chiral clusters clarifies the boundary between forbidden time-crystalline order and allowed current-carrying quantum matter, and highlights a distinct form of many-body organization combining clustering and chirality.

It has been claimed that the ground state of attractive fermions on a ring is rotating \cite{Girardeau2008}. Specifically, this claim concerns the fermionic Tonks-Girardeau gas describing a spin-polarized Fermi gas subject to strongly attractive $p$-wave interactions \cite{GirardeauWright2005, GirardeauMinguzzi2006, MinguzziGirardeau2006}, dual to the ideal Bose gas via the Bose-Fermi mapping \cite{Girardeau60, CheonShigehara99, Girardeau04}. However, such a system exhibits neither clustering nor chirality, as the ground state is doubly degenerate.

Microscopic realizations include attractive one-dimensional Bethe-string clusters known as McGuire quantum solitons \cite{McGuire64, Takahashi99}, whose mean-field behavior describes bright solitons in the Gross-Pitaevskii equation. In a 1D Bose gas \cite{LL63, L63, Olshanii98, LSY03}, steady currents can be associated with an external flux but are otherwise forbidden in the ground state \cite{Kanamoto2009, Kanamoto2010}. By contrast, we show that they are possible in an anyonic attractive Lieb-Liniger gas.

One-dimensional anyons have a recent history. They were first considered as exotic models in the realm of mathematical physics \cite{ZhuWang96, Kundu1999, Batchelor06, Batchelor06PRB, Girardeau2006, Batchelor2007, BatchelorGuan2008}, which nonetheless spurred a wide variety of theoretical efforts exploring their ground state properties \cite{HaoZhangChen08, HaoZhangChen09}, correlations \cite{Calabrese07, Patu07, Santachiara07, Santachiara08, Patu08, Patu08b,Patu09, Patu09b, Patu10, Patu15}, and dynamics \cite{delcampo08, HaoChen12, Piroli17}. 
Following proposals for their quantum simulation \cite{Keilmann11, Longhi12, GreschnerSantos15, Eckardt16, YuanFan17, ZhangGreschner17, Greschner18, LiuGorshkov18, Mackel2023}, their experimental realization with ultracold gases has been reported. In particular, both hard-core anyons \cite{Dhar2025} and Bose-Hubbard anyons \cite{Bakkalihassani2026} have been realized. 

{\it Minimal field theory.---}
To illustrate the concept of chiral cluster, we start by considering a minimal
field theory for a complex scalar field $\psi(\theta,t)$ on a ring. In units $\hbar=2m=1$, used throughout, its dynamics is
governed by the Lagrangian density
\begin{equation}\label{minimal_field_theory}
\mathcal{L}
= i\psi^{\ast}\partial_t\psi
-\left|(\partial_{\theta}-i\alpha\rho)\psi\right|^2
-\frac{g}{2}\rho^2,
\qquad g<0.
\end{equation}
Here, $\theta\in[0,2\pi)$ is the angular coordinate, $\psi$ is
$2\pi$-periodic, and $\rho=|\psi|^2$ is the matter density, normalized to $\int_0^{2\pi}\rho\,d\theta=1$.
The attractive coupling $g$ energetically favors density localization, while the density-dependent vector potential $\alpha\rho$, with $\alpha$ a constant coupling, favors a variation in the phase of $\psi$ around the ring that competes with the periodicity constraint, as we shall see below. This suggests the possibility of a ground state with localized density and a persistent current.
We note that a bosonic model closely related to Eq. (\ref{minimal_field_theory}) was proposed to realize a quantum time crystal as a rotating ground-state soliton \cite{Ohberg2019}. It was later pointed out that the true ground state has a static density profile \cite{Syrwid2020Comment,Syrwid20PRR}, giving rise to a debate \cite{Ohberg2020Reply,Syrwid2021,Ohberg2024,Ohberg2026}.

Under the transformation $\psi(\theta)=
\exp\!\left[
i\alpha\int_0^\theta d\theta'
\left(|\chi(\theta')|^2-\frac1{2\pi}\right)
\right]\chi(\theta)$, the energy functional takes the form
\begin{equation}\label{E_functional_field_theory}
\mathcal{E}=\int_0^{2\pi}d\theta\left[
|(-i\partial_\theta-\Omega)\chi|^2
+\pi\gamma|\chi|^4
\right],
\end{equation}
with $\Omega=\alpha/2\pi$ and $\gamma=g/2\pi$. The minimization of (\ref{E_functional_field_theory}) at fixed norm requires
\begin{equation}\label{GPE_rotating_ring}
    [(-i\partial_{\theta}-\Omega)^{2}+2\pi\gamma|\chi|^{2}]\chi=\mu \chi,
\end{equation}
where $\mu$ is the chemical potential.
The ground-state problem of the theory~(\ref{minimal_field_theory}) is thus equivalent to solving the Gross--Pitaevskii equation on a ring rotating at angular velocity $\Omega$. The solutions of the latter have been obtained analytically and classified \cite{Carr2000, Kanamoto2003}. Specifically,
write $\chi=\sqrt{\rho}e^{i\phi}$. Periodicity requires $\phi(2\pi)-\phi(0)=2\pi J$, with $J\in \mathbb{Z}$, and energy minimization selects $J=\lfloor\Omega+1/2\rfloor$, the integer closest
to $\Omega$. Defining $w=\Omega-J\in[-1/2,1/2)$, one finds that the ground state is a bright soliton for $\gamma<2w^2-1/2$ and uniform otherwise \cite{Kanamoto2003}. In the soliton regime, its density is given in terms of the Jacobi $\operatorname{dn}$ elliptic function by \cite{Kanamoto2003}
\begin{equation}\label{rho_soliton}
\begin{gathered}
\rho_0(\theta)=\frac{K(m)^2}{\pi^3|\gamma|}
\left[\operatorname{dn}^2\!\left(\frac{K(m)}{\pi}(\theta-\theta_0)
\,\middle|\,m\right)-\eta(1-m)\right], \\[4pt]
\eta=\frac{2K(m)E(m)+\pi^2\gamma}{2(1-m)K(m)^2}.
\end{gathered}
\end{equation}
Here, $K(m)$ and $E(m)$ are complete elliptic integrals whose parameter $m$ is fixed by the phase-winding condition above, and $\theta_{0}$ is an arbitrary position shift, reflecting translation symmetry. Further, the conserved particle current is 
\begin{equation}
j = 2\operatorname{Im}[\psi^*(\partial_\theta-i\alpha\rho)\psi] = 2\rho(\partial_{\theta}\phi-\Omega).
\end{equation}
where the second equality follows from the transformation to $\chi$. For the stationary ground state, the continuity equation implies that $j_{0}$ is spatially uniform. Integrating $j_{0}/\rho_{0}=2(\phi'-\Omega)$ around the ring and imposing the phase winding condition yields
\begin{equation}
    j_{0}=-\frac{4\pi w}{\int_{0}^{2\pi}\frac{d\theta}{\rho_{0}(\theta)}}.
\end{equation}
For $0<|w|<1/2$ the soliton density is strictly positive and therefore the current is nonzero.
We have thus shown, at the mean-field level, that the ground state of (\ref{minimal_field_theory}) can exhibit a localized density profile carrying a persistent current, whose direction, set by the sign of $\alpha$, gives the state a handedness. The current is intrinsic, arising from interactions through the density-dependent gauge potential $\alpha \rho$ rather than from a prescribed external vector potential. We refer to such self-bound ground states with an intrinsic persistent current as {\it ground-state chiral quantum clusters}.

{\it Attractive anyonic Lieb--Liniger model.---}
As a paradigmatic example of an exactly solvable system whose ground state forms a chiral quantum cluster, we consider the attractive anyonic Lieb--Liniger model on a ring of length $L$ \cite{LiebLiniger1963,Lieb1963,Kundu1999,Batchelor06,Patu07}. The $N$-particle first-quantized Hamiltonian reads
\begin{equation}\label{H_LL}
    H=-\sum_{j=1}^{N}\partial_{x_{j}}^{2}+2c\sum_{i<j}\delta(x_{i}-x_{j}),
\end{equation}
where $c<0$ is the attractive contact coupling.
The anyonic character of the model is encoded in the commutation relations of the field operators $\hat\Psi_A(x)$,
\begin{equation}\label{anyonic_comm_relations}
\left\{
\begin{aligned}
\hat\Psi_A(x)\hat\Psi_A(y)
&=e^{i\pi\kappa\epsilon(x-y)}\hat\Psi_A(y)\hat\Psi_A(x),\\[4pt]
\hat\Psi_A^\dagger(x)\hat\Psi_A^\dagger(y)
&=e^{i\pi\kappa\epsilon(x-y)}\hat\Psi_A^\dagger(y)\hat\Psi_A^\dagger(x),\\[4pt]
\hat\Psi_A(x)\hat\Psi_A^\dagger(y)
&=e^{-i\pi\kappa\epsilon(x-y)}\hat\Psi_A^\dagger(y)\hat\Psi_A(x)+\delta(x-y).
\end{aligned}
\right.
\end{equation}
Here, $\epsilon$ is the sign function, and $\kappa$ denotes the statistical parameter interpolating between bosonic ($\kappa=0$) and fermionic ($\kappa=1$) exchange. The relations in Eqs. (\ref{anyonic_comm_relations}) are reflected in the transformation properties of the $N$-body wave function $\Psi(x_1,\ldots,x_N)$ under permutations of its arguments. In particular, with the ordering convention $\Psi(x_1,\ldots,x_N)=\langle0|\hat\Psi_A(x_N)\cdots\hat\Psi_A(x_1)|\Phi_N\rangle/\sqrt{N!}$, where $|\Phi_N\rangle$ is an $N$-particle state, interchanging two adjacent arguments yields $\Psi(\ldots,x_i,x_{i+1},\ldots)=e^{-i\pi\kappa\epsilon(x_i-x_{i+1})}\Psi(\ldots,x_{i+1},x_i,\ldots)$.
Factoring out the anyonic exchange phase maps the coordinate eigenvalue problem onto that of the bosonic Lieb--Liniger model with effective coupling $c^{\prime}=c/\cos(\pi \kappa/2)$ \cite{Kundu1999}.
The many-body eigenstates of (\ref{H_LL}) can be constructed via the coordinate Bethe-ansatz as a superposition of plane waves $\Psi(x_{1}, \cdots,x_{N})=\sum_{P} A_{P}\,e^{i\sum_{j=1}^{N}\lambda_{P_{j}}x_{j}}$, where the sum runs over all permutations $P$ of the parameters $\{\lambda_{j}\}$, known as rapidities \cite{Kundu1999, Batchelor06, Patu07}.
Quantizing the theory on the ring consistently with the anyonic exchange relations leads to the Bethe equations \cite{Patu07}
\begin{equation}\label{Bethe_eqs}
    e^{i\lambda_{j}L}=e^{i\pi\kappa (N-1)} \prod_{l\neq j}\frac{\lambda_{j}-\lambda_{l}+ic^{\prime}}{\lambda_{j}-\lambda_{l}-ic^{\prime}}.
\end{equation}
Equation (\ref{Bethe_eqs}) has the same structure as its bosonic counterpart, up to an overall statistical phase. As we shall see, this seemingly simple modification has an important consequence: it allows the ground state to carry nonzero momentum. The solutions of (\ref{Bethe_eqs}) organize into strings of complex rapidities \cite{Patu07, CalabreseCauxPRL2007}. Specifically, in the large $L$ limit, the ground state corresponds to a single $N$-string \cite{CalabreseCauxPRL2007},
\begin{equation}\label{N_cluster_rapidities}
    \lambda^{N,a}=q+i\frac{\bar{c}^{\prime}}{2}(N+1-2a)+i\delta^{N,a}, \quad a=1, \cdots, N,
\end{equation}
where $q$ is the real string center, $\bar c'=-c'>0$, and $\delta^{N,a}$ are finite-size string deviations that vanish exponentially as $L\to\infty$. Equation (\ref{N_cluster_rapidities}) describes an $N$-particle bound state and, up to the phase factor required by the anyonic exchange relations, the corresponding wavefunction is that of a many-body bright quantum soliton \cite{McGuire64}
\begin{equation}
     \Psi_A(x_1,\dots,x_N)\propto
    e^{iq\sum_{j=1}^N x_j}
    \exp\!\left[-\frac{\bar c'}{2}\sum_{i<j}|x_i-x_j|\right],
\end{equation}
with energy $E^{N}=\sum_{a}(\lambda^{N,a})^{2}=Nq^{2}-\frac{\bar{c}'^2}{12}N(N^{2}-1)$ \cite{McGuire64}.
Taking the product of Eq. (\ref{Bethe_eqs}) over all the rapidities yields the quantization of the total momentum, $P=\sum_{j}\lambda_{j}=2\pi n/L+\pi \kappa N(N-1)/L$, with $n\in\mathbb Z$. For an $N$-string, $P=Nq$, so the allowed string centers are $q_n=2\pi n/(NL)+\pi\kappa(N-1)/L$. The ground state minimizes the energy by selecting the value of $q_{n}$ closest to zero. 
For bosons, ($\kappa=0$), this gives $q=0$ and hence $P_{\text{GS}}=0$.
In the anyonic case, by contrast,
\begin{equation}\label{P_GS_eq}
P_{\mathrm{GS}}
=\frac{2\pi n_{\mathrm{GS}}+\pi\kappa N(N-1)}{L},
\end{equation}
where $n_{\text{GS}}$ is the nearest integer to $-\kappa N(N-1)/2$. 
The magnitude of the statistical momentum shift is maximized in the fermionic case, but there, contact interactions are overcome by Pauli exclusion principle.
The resulting ground-state momentum is thus generally nonzero, with $|P_{\mathrm{GS}}|\leq\pi/L$, and particle current $j_{\text{GS}}=2P_{\text{GS}}/L$.
Further, for $q=0$, the mean-field description of such many-body state corresponds to a bright soliton solution of the nonlinear Schr\"odinger equation \cite{CalogeroDegasperis1975,WadatiSakagami1984,CastinHerzog2001}, and thus, its anyonic counterpart describes a boosted bright soliton in the classical limit. The attractive anyonic Lieb--Liniger model therefore provides a simple realization of a self-bound ground-state chiral quantum cluster, in which the bound-state string structure and the anyonic phase account respectively for clustering and chirality.

{\it Dynamical structure factor.---}
As an experimentally accessible diagnostic of chiral clusters, we consider the dynamical structure factor (DSF), that is, the Fourier transform of the density-density correlation function $S(x,t)=\langle \rho(x,t)\rho(0,0)\rangle$:
\begin{equation}\label{DSF_def}
    S(k,w)=\int_{0}^{L} dx \int_{-\infty}^{\infty} dt \, e^{i(wt-kx)}S(x,t),
\end{equation}
which can be measured by Bragg spectroscopy in ultracold atomic gases \cite{Meinert2015} or by neutron scattering in condensed-matter systems \cite{Mourigal2013}.
Inserting a complete set of $N$-particle Bethe eigenstates $\ket{\mu}$, labeled by their rapidities $\{\mu\}$, with energy $E_{\mu}$ and momentum $P_{\mu}$, the ground-state correlator takes the Lehmann form
\begin{equation}\label{S_x_t_lehman}
S(x,t)=\sum_{\mu}\frac{|\Sigma_{\mu, \text{GS}}^{\rho}|^2}{\|\mu\|^{2} \|\text{GS}\|^{2}}\,e^{i(P_{\mu}-P_{\text{GS}})x-i(E_{\mu}-E_{\text{GS}})t},
\end{equation}
where $\|\sigma\|$ is the norm of $\ket{\sigma}$ and  $\Sigma_{\mu, \text{GS}}^{\rho}=\langle \mu|\rho(0,0) |\text{GS}\rangle$ is the density form factor. Thus, Eq. (\ref{DSF_def}) becomes
\begin{equation}\label{S_k_w_lehman}
    S(k,w)=2\pi L \sum_{\mu}\delta(w+E_{\text{GS}}-E_{\mu}) \delta_{k,P_{\mu}-P_{\text{GS}}}\frac{|\Sigma_{\mu, \text{GS}}^{\rho}|^2}{\|\mu\|^{2} \|\text{GS}\|^{2}}.
\end{equation}
The exact expression for $\Sigma^{\rho}_{\mu, \text{GS}}$ has been derived for the bosonic Lieb–Liniger model using the algebraic Bethe ansatz formalism
\cite{Slavnov1989, Slavnov1990,CalabreseCaux2006}.
Leveraging the bosonic-anyonic mapping, the result can be extended to the anyonic case.
Following the conventions of \cite{CalabreseCaux2006}, we write the rapidities $\{\mu\}$ in string form as $\mu_{\alpha}^{j,a}=\mu_{\alpha}^{j}+\frac{i\bar{c}^{\prime}}{2}(j+1-2a),\; a=1,\dots,j$, where $\mu_{\alpha}^{j}$ is the string center, and $\alpha$ labels different strings of same length $j$.
The corresponding anyonic density form factor  reads (see Supplemental Material)
\begin{equation}\label{density_form_factor_anyon}
|\Sigma^{\rho}_{\mu,\text{GS}}|
=\frac{N\,\Gamma^{2}(N)\,\big(P_{\mu}-P_{\text{GS}}\big)^{2}/\bar{c}^{\prime}}{\displaystyle\prod_{j,\alpha,a}
\left[\left(\frac{\mu_{\alpha}^{j}-q}{\bar{c}^{\prime}}\right)^{2}
+\left(\frac{N+j}{2}-a\right)^{2}\right]}\,.
\end{equation}
The full DSF follows by inserting the above expression into Eq. (\ref{S_k_w_lehman}) and evaluating the sum over intermediate states. Consider the contribution from single $N$-string states $\ket{\mu}$ with center $\mu$, which describe the motion of the intact cluster. The momentum constraint in Eq. (\ref{S_k_w_lehman}) fixes $k=N(\mu-q)$ and thus Eq. (\ref{density_form_factor_anyon}) reduces to $|\Sigma^{\rho}_{\mu,\text{GS}}|=\frac{N^{3}\bar{c}^{\prime}}{ \prod_{a=1}^{N-1}\left[(\frac{k}{\bar{c}^{\prime}N a})^{2}+1\right]}$.
Further, for this class of states the excitation energy is $E_\mu-E_{\text{GS}}=k^2/N+2qk$.
Combining these with the $N$-string norms \cite{Piroli2020}, one obtains
\begin{equation}\label{S_N_kw}
    S_{N}(k,w)=\frac{2\pi N^{2}}{L}\frac{\delta(w-\frac{k^2}{N}-2qk)}{\prod_{a=1}^{N-1}\left[\left(\frac{k}{\bar{c}^{\prime}N a}\right)^{2}+1\right]^{2}}.
\end{equation}
This contribution appears in the DSF as a sharp line at $w=k^{2}/N+2qk$, separated from the continuum of states in which the cluster is fragmented by a gap of at least $\bar{c}^{\prime 2} N(N-1)/4$, the energy needed to detach one particle. The isolated line corresponds to the dispersion of a free object of mass $N/2$ moving at velocity $2q$, the signature of clustering.
Further, the term $2qk$ makes the response asymmetric under $k\rightarrow -k$, $S(k,w)\neq S(-k,w)$, and thus reveals the chirality.

{\it Many-body chiral Talbot revivals.---}
The coexistence of clustering and chirality in the ground state of a many-body system is reminiscent of the rotating soliton envisioned in the original time-crystal proposal \cite{Wilczek2012}. It is then natural to ask whether chiral quantum clusters exhibit signatures of such periodic motion. The ground state has definite momentum, so the center of mass of the cluster is delocalized around the ring and the density is uniform, $\langle \rho \rangle=N/L$. Yet, density correlations can reveal the internal structure and chirality of the bound state.
We therefore consider the real-space correlator $S(x,t)$, which is closely related to the van Hove function \cite{VanHove1954} and whose real part can be accessed experimentally through weak measurements \cite{Altuntas2025, Altuntas2026}. 
Inverse Fourier transforming Eq. (\ref{S_N_kw}) yields its $N$-string contribution,
\begin{equation}\label{S_N_x_t}
    S_{N}(x,t)=\frac{N^{2}}{L^{2}}\sum_{k}\frac{e^{ik(x-2qt)-ik^{2}t/N}}{\prod_{a=1}^{N-1}\left[\left(\frac{k}{\bar{c}^{\prime} N a}\right)^{2}+1\right]^{2}},
\end{equation}
that is, the sum in Eq. (\ref{S_x_t_lehman}) restricted to $N$-string intermediate $\ket{\mu}$ states.
\begin{figure}
    \centering
    \includegraphics[width=0.8
\linewidth]{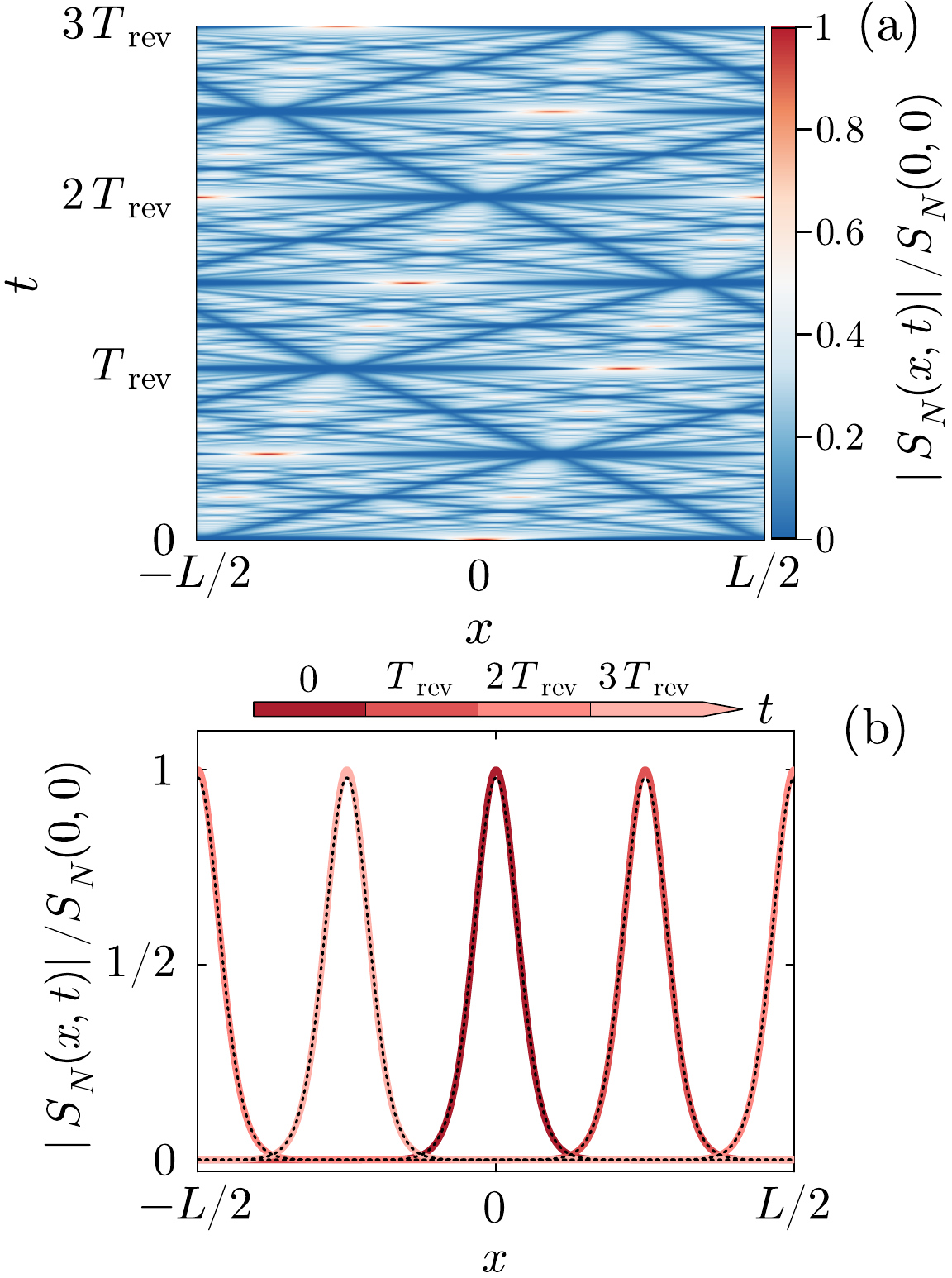}
    \caption{Signature of a ground-state chiral quantum cluster in the density-density correlator. (a) Quantum carpet generated by the $N$-string contribution $S_{N}(x,t)$ to the density-density correlator, Eq. (\ref{S_N_x_t}). (b) $S_{N}(x,t)$ at integer multiples of the revival time $T_{\mathrm{rev}}$. Shades of red correspond to the direct summation of Eq. (\ref{S_N_x_t}); dashed black lines show the closed form of Eq. (\ref{N_string_bump_profile}) shifted by $2qt$. The parameter values used in both panels are $N=20$, $L=10\pi$, $\bar c' = 0.1$, and $\kappa N(N-1) = 190 + 1/4$.}
    \label{fig:quantum_carpet_and_revivals}
\end{figure}
At $t=0$ and large $N$, Eq. (\ref{S_N_x_t}) can be evaluated in closed form (see Supplemental Material), giving to leading order
\begin{equation}\label{N_string_bump_profile}
    S_{N}(x,0)\simeq \frac{\cbar^{\prime}N^{3}}{2L} \frac{u\cosh{u}-\sinh{u}}{\sinh^{3}{u}}, \quad u=\frac{\cbar^{\prime} N x}{2},
\end{equation} 
which describes a localized bump of characteristic width $\xi=1/(\cbar^{\prime}N)$, reflecting the clustering of particles in the bound state.
The remaining intermediate states in Eq.~(\ref{S_x_t_lehman}) describe a fragmented cluster. Writing $\rho(x)=\sum_i\delta(x-x_i)$, it follows that
$S(x,0)=(N/L)\,\delta(x)+\big\langle\textstyle\sum_{i\neq j}\delta(x-x_i)\,\delta(x_j)\big\rangle$.
The first term is the self contribution, on which we comment further below, while the second is the regular part of the correlator, fixed at $x=0$ by the Hellmann--Feynman theorem to $\cbar^{\prime}N(N^{2}-1)/6L$ \cite{CalabreseCaux2006}, whose leading term is saturated by $S_{N}(0,0)\simeq\cbar^{\prime}N^{3}/6L$ from Eq.~(\ref{N_string_bump_profile}).
Experimentally, the finite spatial resolution of density measurements
introduces a momentum cutoff $|k|<k_{\max}$ in Eq.~(\ref{S_x_t_lehman}) \cite{Altuntas2025,Altuntas2026}.
Resolving the profile of Eq. (\ref{N_string_bump_profile}) requires $k_{\max}\gg1/\xi$.
Further, for $k_{\max}\ll N/\xi$, the self term is regularized into a peak of height $\sim Nk_{\max}/L$, which is itself of order $1/N$ relative to $S_N(0,0)$. Since all weights in Eq.~(\ref{S_x_t_lehman}) are positive, the contribution of the fragmented states then remains of relative order $1/N$ with respect to
$S_N(0,0)$ at any $x$ and $t$. 
With increasing time separation, the profile initially spreads as its Fourier components dephase.
However, crucially, the momentum transfer $k=P_{\mu}-P_{\text{GS}}$ is quantized as $2\pi n/L$. The quadratic phase factors in Eq. (\ref{S_N_x_t}) therefore return to unity simultaneously at integer multiples of the full revival time
\begin{equation}\label{eq:T_rev}
    T_{\mathrm{rev}}=\frac{NL^2}{2\pi},
\end{equation}
since $e^{-ik_n^2T_{\mathrm{rev}}/N}=e^{-2\pi i n^2}=1$ for every $n$.
The original ($t=0$) shape of the bump is thus restored at these times. This recurrence is a manifestation of the Talbot effect \cite{Mack2002,Marzoli1998}, shown in Fig. \ref{fig:quantum_carpet_and_revivals}(a). Moreover, the correlation pattern drifts at velocity $2q$, thus the reconstructed bump reappears shifted by $2qT_{\mathrm{rev}}$ from its position at the preceding revival (Fig. \ref{fig:quantum_carpet_and_revivals}(b)). Its successive positions trace a motion around the ring whose direction reflects the chirality of the ground state ($q=P_{\text{GS}}/N$).

{\it Conclusion.---}
The idea of spontaneous breaking of time translation symmetry opened up the fascinating possibility of observing a periodic motion in the lowest energy state of a quantum system at equilibrium, such as the persistent rotation of a ground-state soliton along a ring \cite{Wilczek2012}.
Motivated by this picture, we have described ground-state chiral quantum clusters: equilibrium many-body ground states exhibiting strong clustering in density correlations and carrying an intrinsic persistent current. The anyonic Lieb--Liniger model on a ring offers an exactly solvable realization: by means of the Bethe ansatz, its ground state is found to be a McGuire bright soliton, displaying a persistent current set by the anyonic statistical parameter.
Strikingly, the ground-state density-density correlator at unequal times and positions exhibits the Talbot effect, with a soliton-like profile that disperses and revives periodically as it drifts along the ring.
This behavior corresponds to a mesoscopic quantum time crystal, realized in the ground state, i.e., without external driving or dissipation.  
The dynamical structure factor provides an experimentally feasible diagnostic of chiral quantum clusters, resolving both the center-of-mass recoil of the bound object and the handedness of the persistent current.

{\it Acknowledgments.---} It is a pleasure to acknowledge valuable discussions with Bogdan Damski, Preethi Gopalakrishnan, András Grabarits, Nilanjan Sasmal and Jing Yang. We acknowledge financial support from the Luxembourg National Research Fund under Grant No. C22/MS/17132060/BeyondKZM.  AdC acknowledges the Basque Government’s BERC and the Severo Ochoa Centre of Excellence Programmes (CEX2024-001491-S, funded by MICIU/AEI/10.13039/501100011033).

\renewcommand{\bibsection}{}

\bibliographystyle{apsrev4-2}
\bibliography{ChiralClusters}

@article{LiebLiniger1963,
  author    = {Lieb, Elliott H. and Liniger, Werner},
  title     = {Exact Analysis of an Interacting Bose Gas. {I}. The General Solution and the Ground State},
  journal   = {Phys. Rev.},
  volume    = {130},
  issue     = {4},
  pages     = {1605--1616},
  year      = {1963},
  month     = {May},
  publisher = {American Physical Society},
  doi       = {10.1103/PhysRev.130.1605},
  url       = {https://link.aps.org/doi/10.1103/PhysRev.130.1605}
}

@article{Lieb1963,
  author    = {Lieb, Elliott H.},
  title     = {Exact Analysis of an Interacting Bose Gas. {II}. The Excitation Spectrum},
  journal   = {Phys. Rev.},
  volume    = {130},
  issue     = {4},
  pages     = {1616--1624},
  year      = {1963},
  month     = {May},
  publisher = {American Physical Society},
  doi       = {10.1103/PhysRev.130.1616},
  url       = {https://link.aps.org/doi/10.1103/PhysRev.130.1616}
}

@article{CalogeroDegasperis1975,
  title = {Comparison between the exact and Hartree solutions of a one-dimensional many-body problem},
  author = {Calogero, F. and Degasperis, A.},
  journal = {Phys. Rev. A},
  volume = {11},
  issue = {1},
  pages = {265--269},
  numpages = {0},
  year = {1975},
  month = {Jan},
  publisher = {American Physical Society},
  doi = {10.1103/PhysRevA.11.265},
  url = {https://link.aps.org/doi/10.1103/PhysRevA.11.265}
}

@article{Sacha2018,
doi = {10.1088/1361-6633/aa8b38},
url = {https://doi.org/10.1088/1361-6633/aa8b38},
year = {2017},
month = {nov},
publisher = {IOP Publishing},
volume = {81},
number = {1},
pages = {016401},
author = {Sacha, Krzysztof and Zakrzewski, Jakub},
title = {Time crystals: a review},
journal = {Reports on Progress in Physics}
}

@book{Sacha2020,
  author    = {Sacha, Krzysztof},
  title     = {Time Crystals},
  series    = {Springer Series on Atomic, Optical, and Plasma Physics},
  volume    = {114},
  publisher = {Springer},
  address   = {Cham},
  year      = {2020},
  edition   = {1},
  isbn      = {978-3-030-52522-4},
  doi       = {10.1007/978-3-030-52523-1}
}

@book{Gaudin2014,
  author    = {Gaudin, Michel},
  title     = {The {B}ethe Wavefunction},
  translator = {Caux, Jean-S{\'e}bastien},
  publisher = {Cambridge University Press},
  address   = {Cambridge},
  year      = {2014},
  isbn      = {9781107045859},
  doi       = {10.1017/CBO9781107053885}
}

@article{Korepin1982,
  author  = {Korepin, Vladimir E.},
  title   = {Calculation of norms of {B}ethe wave functions},
  journal = {Communications in Mathematical Physics},
  volume  = {86},
  number  = {3},
  pages   = {391--418},
  year    = {1982},
  doi     = {10.1007/BF01212176}
}

@article{WadatiSakagami1984,
  author  = {Wadati, Miki and Sakagami, Masa},
  title   = {Classical Soliton as a Limit of the Quantum Field Theory},
  journal = {Journal of the Physical Society of Japan},
  volume  = {53},
  number  = {6},
  pages   = {1933--1938},
  year    = {1984},
  doi     = {10.1143/JPSJ.53.1933}
}

@article{CastinHerzog2001,
  author        = {Castin, Yvan and Herzog, Christopher},
  title         = {{B}ose--{E}instein condensates in symmetry breaking states},
  journal       = {Comptes Rendus de l'Acad{\'e}mie des Sciences - Series IV - Physics},
  volume        = {2},
  number        = {3},
  pages         = {419--443},
  year          = {2001},
  doi           = {10.1016/S1296-2147(01)01183-0}
}

@article{Kozin2019,
  title = {Quantum Time Crystals from Hamiltonians with Long-Range Interactions},
  author = {Kozin, Valerii K. and Kyriienko, Oleksandr},
  journal = {Phys. Rev. Lett.},
  volume = {123},
  issue = {21},
  pages = {210602},
  numpages = {6},
  year = {2019},
  month = {Nov},
  publisher = {American Physical Society},
  doi = {10.1103/PhysRevLett.123.210602},
  url = {https://link.aps.org/doi/10.1103/PhysRevLett.123.210602}
}

@article{Zaletel2023,
  title = {Colloquium: Quantum and classical discrete time crystals},
  author = {Zaletel, Michael P. and Lukin, Mikhail and Monroe, Christopher and Nayak, Chetan and Wilczek, Frank and Yao, Norman Y.},
  journal = {Rev. Mod. Phys.},
  volume = {95},
  issue = {3},
  pages = {031001},
  numpages = {34},
  year = {2023},
  month = {Jul},
  publisher = {American Physical Society},
  doi = {10.1103/RevModPhys.95.031001},
  url = {https://link.aps.org/doi/10.1103/RevModPhys.95.031001}
}

@misc{Bakkalihassani2026,
      title={Revealing Pseudo-Fermionization and Chiral Binding of One-Dimensional Anyons using Adiabatic State Preparation}, 
      author={Brice Bakkali-Hassani and Joyce Kwan and Perrin Segura and Yanfei Li and Isaac Tesfaye and Gerard Valentí-Rojas and André Eckardt and Markus Greiner},
      year={2026},
      eprint={2602.20421},
      archivePrefix={arXiv},
      primaryClass={cond-mat.quant-gas},
      url={https://arxiv.org/abs/2602.20421}, 
}

@Article{Dhar2025,
author={Dhar, Sudipta
and Wang, Botao
and Horvath, Milena
and Vashisht, Amit
and Zeng, Yi
and Zvonarev, Mikhail B.
and Goldman, Nathan
and Guo, Yanliang
and Landini, Manuele
and N{\"a}gerl, Hanns-Christoph},
title={Observing anyonization of bosons in a quantum gas},
journal={Nature},
year={2025},
month={Jun},
day={01},
volume={642},
number={8066},
pages={53-57},
issn={1476-4687},
doi={10.1038/s41586-025-09016-9},
url={https://doi.org/10.1038/s41586-025-09016-9}
}

@article{Calabrese07,
	author = {Calabrese, Pasquale and Mintchev, Mihail},
	doi = {10.1103/PhysRevB.75.233104},
	issue = {23},
	journal = {Phys. Rev. B},
	month = {Jun},
	numpages = {4},
	pages = {233104},
	publisher = {American Physical Society},
	title = {Correlation functions of one-dimensional anyonic fluids},
	url = {https://link.aps.org/doi/10.1103/PhysRevB.75.233104},
	volume = {75},
	year = {2007}}

@article{Patu07,
	author = {Ovidiu I P{\^{a}}{\c{t}}u and Vladimir E Korepin and Dmitri V Averin},
	doi = {10.1088/1751-8113/40/50/004},
	journal = {Journal of Physics A: Mathematical and Theoretical},
	month = {nov},
	number = {50},
	pages = {14963--14984},
	publisher = {{IOP} Publishing},
	title = {Correlation functions of one-dimensional Lieb{\textendash}Liniger anyons},
	url = {https://doi.org/10.1088/1751-8113/40/50/004},
	volume = {40},
	year = 2007}

@article{Santachiara08,
	author = {Raoul Santachiara and Pasquale Calabrese},
	doi = {10.1088/1742-5468/2008/06/p06005},
	journal = {Journal of Statistical Mechanics: Theory and Experiment},
	month = {jun},
	number = {06},
	pages = {P06005},
	publisher = {{IOP} Publishing},
	title = {One-particle density matrix and momentum distribution function of one-dimensional anyon gases},
	url = {https://doi.org/10.1088/1742-5468/2008/06/p06005},
	volume = {2008},
	year = 2008}

@article{ZhuWang96,
	author = {Zhu, Jian-Xin and Wang, Z. D.},
	doi = {10.1103/PhysRevA.53.600},
	issue = {1},
	journal = {Phys. Rev. A},
	month = {Jan},
	numpages = {0},
	pages = {600--603},
	publisher = {American Physical Society},
	title = {Topological effects associated with fractional statistics in one-dimensional mesoscopic rings},
	url = {https://link.aps.org/doi/10.1103/PhysRevA.53.600},
	volume = {53},
	year = {1996}}

@article{Santachiara07,
	author = {Raoul Santachiara and Franck Stauffer and Daniel C Cabra},
	doi = {10.1088/1742-5468/2007/05/l05003},
	journal = {Journal of Statistical Mechanics: Theory and Experiment},
	month = {may},
	number = {05},
	pages = {L05003--L05003},
	publisher = {{IOP} Publishing},
	title = {Entanglement properties and momentum distributions of hard-core anyons on a ring},
	url = {https://doi.org/10.1088/1742-5468/2007/05/l05003},
	volume = {2007},
	year = 2007}

@article{HaoZhangChen08,
	author = {Hao, Yajiang and Zhang, Yunbo and Chen, Shu},
	doi = {10.1103/PhysRevA.78.023631},
	issue = {2},
	journal = {Phys. Rev. A},
	month = {Aug},
	numpages = {6},
	pages = {023631},
	publisher = {American Physical Society},
	title = {Ground-state properties of one-dimensional anyon gases},
	url = {https://link.aps.org/doi/10.1103/PhysRevA.78.023631},
	volume = {78},
	year = {2008}}

@article{HaoZhangChen09,
	author = {Hao, Yajiang and Zhang, Yunbo and Chen, Shu},
	doi = {10.1103/PhysRevA.79.043633},
	issue = {4},
	journal = {Phys. Rev. A},
	month = {Apr},
	numpages = {5},
	pages = {043633},
	publisher = {American Physical Society},
	title = {Ground-state properties of hard-core anyons in one-dimensional optical lattices},
	url = {https://link.aps.org/doi/10.1103/PhysRevA.79.043633},
	volume = {79},
	year = {2009}}

@article{HaoChen12,
	author = {Hao, Yajiang and Chen, Shu},
	doi = {10.1103/PhysRevA.86.043631},
	issue = {4},
	journal = {Phys. Rev. A},
	month = {Oct},
	numpages = {6},
	pages = {043631},
	publisher = {American Physical Society},
	title = {Dynamical properties of hard-core anyons in one-dimensional optical lattices},
	url = {https://link.aps.org/doi/10.1103/PhysRevA.86.043631},
	volume = {86},
	year = {2012}}

@article{Patu08,
	author = {Ovidiu I P{\^{a}}{\c{t}}u and Vladimir E Korepin and Dmitri V Averin},
	doi = {10.1088/1751-8113/41/14/145006},
	journal = {Journal of Physics A: Mathematical and Theoretical},
	month = {mar},
	number = {14},
	pages = {145006},
	publisher = {{IOP} Publishing},
	title = {One-dimensional impenetrable anyons in thermal equilibrium: I. Anyonic generalization of Lenard's formula},
	url = {https://doi.org/10.1088/1751-8113/41/14/145006},
	volume = {41},
	year = 2008}

@article{Patu08b,
	author = {Ovidiu I P{\^{a}}{\c{t}}u and Vladimir E Korepin and Dmitri V Averin},
	doi = {10.1088/1751-8113/41/25/255205},
	journal = {Journal of Physics A: Mathematical and Theoretical},
	month = {may},
	number = {25},
	pages = {255205},
	publisher = {{IOP} Publishing},
	title = {One-dimensional impenetrable anyons in thermal equilibrium: {II}. Determinant representation for the dynamic correlation functions},
	url = {https://doi.org/10.1088/1751-8113/41/25/255205},
	volume = {41},
	year = 2008}

@article{Patu09,
	author = {O. I. P{\^{a}}{\c{t}}u and V. E. Korepin and D. V. Averin},
	doi = {10.1209/0295-5075/86/40001},
	journal = {{EPL} (Europhysics Letters)},
	month = {may},
	number = {4},
	pages = {40001},
	publisher = {{IOP} Publishing},
	title = {Large-distance asymptotic behavior of the correlation functions of 1D impenetrable anyons at finite temperatures},
	url = {https://doi.org/10.1209/0295-5075/86/40001},
	volume = {86},
	year = 2009}

@article{Patu09b,
	author = {Ovidiu I P{\^{a}}{\c{t}}u and Vladimir E Korepin and Dmitri V Averin},
	doi = {10.1088/1751-8113/42/27/275207},
	journal = {Journal of Physics A: Mathematical and Theoretical},
	month = {jun},
	number = {27},
	pages = {275207},
	publisher = {{IOP} Publishing},
	title = {One-dimensional impenetrable anyons in thermal equilibrium: {III}. Large distance asymptotics of the space correlations},
	url = {https://doi.org/10.1088/1751-8113/42/27/275207},
	volume = {42},
	year = 2009}

@article{Patu10,
	author = {Ovidiu I P{\^{a}}{\c{t}}u and Vladimir E Korepin and Dmitri V Averin},
	doi = {10.1088/1751-8113/43/11/115204},
	journal = {Journal of Physics A: Mathematical and Theoretical},
	month = {mar},
	number = {11},
	pages = {115204},
	publisher = {{IOP} Publishing},
	title = {One-dimensional impenetrable anyons in thermal equilibrium: {IV}. Large time and distance asymptotic behavior of the correlation functions},
	url = {https://doi.org/10.1088/1751-8113/43/11/115204},
	volume = {43},
	year = 2010}

@article{Patu15,
	author = {Ovidiu I P{\^{a}}{\c{t}}u},
	doi = {10.1088/1742-5468/2015/01/p01004},
	journal = {Journal of Statistical Mechanics: Theory and Experiment},
	month = {jan},
	number = {1},
	pages = {P01004},
	publisher = {{IOP} Publishing},
	title = {Correlation functions and momentum distribution of one-dimensional hard-core anyons in optical lattices},
	url = {https://doi.org/10.1088/1742-5468/2015/01/p01004},
	volume = {2015},
	year = 2015}

@article{delcampo08,
	author = {del Campo, A.},
	doi = {10.1103/PhysRevA.78.045602},
	issue = {4},
	journal = {Phys. Rev. A},
	month = {Oct},
	numpages = {4},
	pages = {045602},
	publisher = {American Physical Society},
	title = {Fermionization and bosonization of expanding one-dimensional anyonic fluids},
	url = {https://link.aps.org/doi/10.1103/PhysRevA.78.045602},
	volume = {78},
	year = {2008}}

@article{Piroli17,
	author = {Piroli, Lorenzo and Calabrese, Pasquale},
	doi = {10.1103/PhysRevA.96.023611},
	issue = {2},
	journal = {Phys. Rev. A},
	month = {Aug},
	numpages = {22},
	pages = {023611},
	publisher = {American Physical Society},
	title = {Exact dynamics following an interaction quench in a one-dimensional anyonic gas},
	url = {https://link.aps.org/doi/10.1103/PhysRevA.96.023611},
	volume = {96},
	year = {2017}}

@article{LiuGorshkov18,
	author = {Liu, Fangli and Garrison, James R. and Deng, Dong-Ling and Gong, Zhe-Xuan and Gorshkov, Alexey V.},
	doi = {10.1103/PhysRevLett.121.250404},
	issue = {25},
	journal = {Phys. Rev. Lett.},
	month = {Dec},
	numpages = {7},
	pages = {250404},
	publisher = {American Physical Society},
	title = {Asymmetric Particle Transport and Light-Cone Dynamics Induced by Anyonic Statistics},
	url = {https://link.aps.org/doi/10.1103/PhysRevLett.121.250404},
	volume = {121},
	year = {2018}}

@article{Batchelor06PRB,
  title = {Generalized exclusion statistics and degenerate signature of strongly interacting anyons},
  author = {Batchelor, M. T. and Guan, X.-W.},
  journal = {Phys. Rev. B},
  volume = {74},
  issue = {19},
  pages = {195121},
  numpages = {7},
  year = {2006},
  month = {Nov},
  publisher = {American Physical Society},
  doi = {10.1103/PhysRevB.74.195121},
  url = {https://link.aps.org/doi/10.1103/PhysRevB.74.195121}
}

@article{Mackel2023,
  doi = {10.22331/q-2023-12-20-1211},
  url = {https://doi.org/10.22331/q-2023-12-20-1211},
  title = {Quantum {A}lchemy and {U}niversal {O}rthogonality {C}atastrophe in {O}ne-{D}imensional {A}nyons},
  author = {Mackel, Naim E. and Yang, Jing and Campo, Adolfo del},
  journal = {{Quantum}},
  issn = {2521-327X},
  publisher = {{Verein zur F{\"{o}}rderung des Open Access Publizierens in den Quantenwissenschaften}},
  volume = {7},
  pages = {1211},
  month = dec,
  year = {2023}
}

@article{Batchelor06,
	author = {Batchelor, M. T. and Guan, X.-W. and Oelkers, N.},
	doi = {10.1103/PhysRevLett.96.210402},
	issue = {21},
	journal = {Phys. Rev. Lett.},
	month = {Jun},
	numpages = {4},
	pages = {210402},
	publisher = {American Physical Society},
	title = {One-Dimensional Interacting Anyon Gas: Low-Energy Properties and Haldane Exclusion Statistics},
	url = {https://link.aps.org/doi/10.1103/PhysRevLett.96.210402},
	volume = {96},
	year = {2006}}

@article{Keilmann11,
	author = {Keilmann, Tassilo and Lanzmich, Simon and McCulloch, Ian and Roncaglia, Marco},
	day = {21},
	doi = {10.1038/ncomms1353},
	issn = {2041-1723},
	journal = {Nature Communications},
	month = {Jun},
	number = {1},
	pages = {361},
	title = {Statistically induced phase transitions and anyons in 1D optical lattices},
	url = {https://doi.org/10.1038/ncomms1353},
	volume = {2},
	year = {2011}}

@article{Girardeau2008,
  title = {Rotating Ground States of a One-Dimensional Spin-Polarized Gas of Fermionic Atoms with Attractive $p$-Wave Interactions on a Mesoscopic Ring},
  author = {Girardeau, M. D. and Wright, E. M.},
  journal = {Phys. Rev. Lett.},
  volume = {100},
  issue = {20},
  pages = {200403},
  numpages = {4},
  year = {2008},
  month = {May},
  publisher = {American Physical Society},
  doi = {10.1103/PhysRevLett.100.200403},
  url = {https://link.aps.org/doi/10.1103/PhysRevLett.100.200403}
}

@article{Girardeau60,
author = {Girardeau, M. },
title = {Relationship between Systems of Impenetrable Bosons and Fermions in One Dimension},
journal = {Journal of Mathematical Physics},
volume = {1},
number = {6},
pages = {516-523},
year = {1960},
doi = {10.1063/1.1703687},
URL = {https://doi.org/10.1063/1.1703687}
}

@article{Girardeau04,
title = "Effective interactions, Fermi–Bose duality, and ground states of ultracold atomic vapors in tight de Broglie waveguides",
journal = "Optics Communications",
volume = "243",
number = "1",
pages = "3 - 22",
year = "2004",
issn = "0030-4018",
doi = "https://doi.org/10.1016/j.optcom.2004.09.079",
url = "http://www.sciencedirect.com/science/article/pii/S0030401804010582",
author = "M.D. Girardeau and Hieu Nguyen and M. Olshanii"
}

@article{Girardeau2006,
  title = {Anyon-Fermion Mapping and Applications to Ultracold Gases in Tight Waveguides},
  author = {Girardeau, M. D.},
  journal = {Phys. Rev. Lett.},
  volume = {97},
  issue = {10},
  pages = {100402},
  numpages = {4},
  year = {2006},
  month = {Sep},
  publisher = {American Physical Society},
  doi = {10.1103/PhysRevLett.97.100402},
  url = {https://link.aps.org/doi/10.1103/PhysRevLett.97.100402}
}

@article{Kundu1999,
  title = {Exact Solution of Double $\ensuremath{\delta}$ Function Bose Gas through an Interacting Anyon Gas},
  author = {Kundu, Anjan},
  journal = {Phys. Rev. Lett.},
  volume = {83},
  issue = {7},
  pages = {1275--1278},
  numpages = {0},
  year = {1999},
  month = {Aug},
  publisher = {American Physical Society},
  doi = {10.1103/PhysRevLett.83.1275},
  url = {https://link.aps.org/doi/10.1103/PhysRevLett.83.1275}
}

@article{GirardeauWright2005,
  title = {Static and Dynamic Properties of Trapped Fermionic Tonks-Girardeau Gases},
  author = {Girardeau, M. D. and Wright, E. M.},
  journal = {Phys. Rev. Lett.},
  volume = {95},
  issue = {1},
  pages = {010406},
  numpages = {4},
  year = {2005},
  month = {Jun},
  publisher = {American Physical Society},
  doi = {10.1103/PhysRevLett.95.010406},
  url = {https://link.aps.org/doi/10.1103/PhysRevLett.95.010406}
}

@article{GirardeauMinguzzi2006,
  title = {Bosonization, Pairing, and Superconductivity of the Fermionic Tonks-Girardeau Gas},
  author = {Girardeau, M. D. and Minguzzi, A.},
  journal = {Phys. Rev. Lett.},
  volume = {96},
  issue = {8},
  pages = {080404},
  numpages = {4},
  year = {2006},
  month = {Mar},
  publisher = {American Physical Society},
  doi = {10.1103/PhysRevLett.96.080404},
  url = {https://link.aps.org/doi/10.1103/PhysRevLett.96.080404}
}

@article{MinguzziGirardeau2006,
  title = {Pairing of a harmonically trapped fermionic Tonks-Girardeau gas},
  author = {Minguzzi, A. and Girardeau, M. D.},
  journal = {Phys. Rev. A},
  volume = {73},
  issue = {6},
  pages = {063614},
  numpages = {4},
  year = {2006},
  month = {Jun},
  publisher = {American Physical Society},
  doi = {10.1103/PhysRevA.73.063614},
  url = {https://link.aps.org/doi/10.1103/PhysRevA.73.063614}
}

@article{CheonShigehara99,
  title = {Fermion-Boson Duality of One-Dimensional Quantum Particles with Generalized Contact Interactions},
  author = {Cheon, Taksu and Shigehara, T.},
  journal = {Phys. Rev. Lett.},
  volume = {82},
  issue = {12},
  pages = {2536--2539},
  numpages = {0},
  year = {1999},
  month = {Mar},
  publisher = {American Physical Society},
  doi = {10.1103/PhysRevLett.82.2536},
  url = {https://link.aps.org/doi/10.1103/PhysRevLett.82.2536}
}

@article{LL63,
  title = {Exact Analysis of an Interacting Bose Gas. I. The General Solution and the Ground State},
  author = {Lieb, Elliott H. and Liniger, Werner},
  journal = {Phys. Rev.},
  volume = {130},
  issue = {4},
  pages = {1605--1616},
  numpages = {0},
  year = {1963},
  month = {May},
  publisher = {American Physical Society},
  doi = {10.1103/PhysRev.130.1605},
  url = {https://link.aps.org/doi/10.1103/PhysRev.130.1605}
}

@article{L63,
  title = {Exact Analysis of an Interacting Bose Gas. II. The Excitation Spectrum},
  author = {Lieb, Elliott H.},
  journal = {Phys. Rev.},
  volume = {130},
  issue = {4},
  pages = {1616--1624},
  numpages = {0},
  year = {1963},
  month = {May},
  publisher = {American Physical Society},
  doi = {10.1103/PhysRev.130.1616},
  url = {https://link.aps.org/doi/10.1103/PhysRev.130.1616}
}

@article{Olshanii98,
  title = {Atomic Scattering in the Presence of an External Confinement and a Gas of Impenetrable Bosons},
  author = {Olshanii, M.},
  journal = {Phys. Rev. Lett.},
  volume = {81},
  issue = {5},
  pages = {938--941},
  numpages = {0},
  year = {1998},
  month = {Aug},
  publisher = {American Physical Society},
  doi = {10.1103/PhysRevLett.81.938},
  url = {https://link.aps.org/doi/10.1103/PhysRevLett.81.938}
}

@article{LSY03,
  title = {One-Dimensional Bosons in Three-Dimensional Traps},
  author = {Lieb, Elliott H. and Seiringer, Robert and Yngvason, Jakob},
  journal = {Phys. Rev. Lett.},
  volume = {91},
  issue = {15},
  pages = {150401},
  numpages = {4},
  year = {2003},
  month = {Oct},
  publisher = {American Physical Society},
  doi = {10.1103/PhysRevLett.91.150401},
  url = {https://link.aps.org/doi/10.1103/PhysRevLett.91.150401}
}

@article{McGuire64,
author = {McGuire,J. B. },
title = {Study of Exactly Soluble One‐Dimensional N‐Body Problems},
journal = {Journal of Mathematical Physics},
volume = {5},
number = {5},
pages = {622-636},
year = {1964},
doi = {10.1063/1.1704156},
URL = {https://doi.org/10.1063/1.1704156}
}

@book{Takahashi99,
  title={Thermodynamics of One-Dimensional Solvable Models},
  author={M. Takahashi},
  year={1999},
  publisher={Cambridge, Cambridge},
  doi={https://doi.org/10.1017/CBO9780511524332}
}

@article{Kanamoto2009,
  title = {Metastable quantum phase transitions in a periodic one-dimensional Bose gas: Mean-field and Bogoliubov analyses},
  author = {Kanamoto, R. and Carr, L. D. and Ueda, M.},
  journal = {Phys. Rev. A},
  volume = {79},
  issue = {6},
  pages = {063616},
  numpages = {12},
  year = {2009},
  month = {Jun},
  publisher = {American Physical Society},
  doi = {10.1103/PhysRevA.79.063616},
  url = {https://link.aps.org/doi/10.1103/PhysRevA.79.063616}
}

@article{Kanamoto2010,
  title = {Metastable quantum phase transitions in a periodic one-dimensional Bose gas. II. Many-body theory},
  author = {Kanamoto, R. and Carr, L. D. and Ueda, M.},
  journal = {Phys. Rev. A},
  volume = {81},
  issue = {2},
  pages = {023625},
  numpages = {13},
  year = {2010},
  month = {Feb},
  publisher = {American Physical Society},
  doi = {10.1103/PhysRevA.81.023625},
  url = {https://link.aps.org/doi/10.1103/PhysRevA.81.023625}
}

@article{Carr2000,
  title = {Stationary solutions of the one-dimensional nonlinear Schr\"odinger equation. II. Case of attractive nonlinearity},
  author = {Carr, L. D. and Clark, Charles W. and Reinhardt, W. P.},
  journal = {Phys. Rev. A},
  volume = {62},
  issue = {6},
  pages = {063611},
  numpages = {10},
  year = {2000},
  month = {Nov},
  publisher = {American Physical Society},
  doi = {10.1103/PhysRevA.62.063611},
  url = {https://link.aps.org/doi/10.1103/PhysRevA.62.063611}
}

@article{CalabreseCauxPRL2007,
  title = {Correlation Functions of the One-Dimensional Attractive Bose Gas},
  author = {Calabrese, Pasquale and Caux, Jean-S{\'e}bastien},
  journal = {Phys. Rev. Lett.},
  volume = {98},
  issue = {15},
  pages = {150403},
  numpages = {4},
  year = {2007},
  month = {Apr},
  publisher = {American Physical Society},
  doi = {10.1103/PhysRevLett.98.150403},
  url = {https://link.aps.org/doi/10.1103/PhysRevLett.98.150403}
}

@article{CalabreseCaux2006,
  author        = {Calabrese, Pasquale and Caux, Jean-S{\'e}bastien},
  title         = {Dynamics of the attractive {1D} {B}ose gas: analytical treatment from integrability},
  journal       = {Journal of Statistical Mechanics: Theory and Experiment},
  volume        = {2007},
  number        = {08},
  pages         = {P08032},
  year          = {2007},
  doi           = {10.1088/1742-5468/2007/08/P08032}
}

@article{Wilczek2012,
  title = {Quantum Time Crystals},
  author = {Wilczek, Frank},
  journal = {Phys. Rev. Lett.},
  volume = {109},
  issue = {16},
  pages = {160401},
  numpages = {5},
  year = {2012},
  month = {Oct},
  publisher = {American Physical Society},
  doi = {10.1103/PhysRevLett.109.160401},
  url = {https://link.aps.org/doi/10.1103/PhysRevLett.109.160401}
}

@article{Bruno2013,
  title = {Impossibility of Spontaneously Rotating Time Crystals: A No-Go Theorem},
  author = {Bruno, Patrick},
  journal = {Phys. Rev. Lett.},
  volume = {111},
  issue = {7},
  pages = {070402},
  numpages = {5},
  year = {2013},
  month = {Aug},
  publisher = {American Physical Society},
  doi = {10.1103/PhysRevLett.111.070402},
  url = {https://link.aps.org/doi/10.1103/PhysRevLett.111.070402}
}

@article{WatanabeOshikawa2015,
  title = {Absence of Quantum Time Crystals},
  author = {Watanabe, Haruki and Oshikawa, Masaki},
  journal = {Phys. Rev. Lett.},
  volume = {114},
  issue = {25},
  pages = {251603},
  numpages = {5},
  year = {2015},
  month = {Jun},
  publisher = {American Physical Society},
  doi = {10.1103/PhysRevLett.114.251603},
  url = {https://link.aps.org/doi/10.1103/PhysRevLett.114.251603}
}

@article{Else2016,
  title = {Floquet Time Crystals},
  author = {Else, Dominic V. and Bauer, Bela and Nayak, Chetan},
  journal = {Phys. Rev. Lett.},
  volume = {117},
  issue = {9},
  pages = {090402},
  numpages = {5},
  year = {2016},
  month = {Aug},
  publisher = {American Physical Society},
  doi = {10.1103/PhysRevLett.117.090402},
  url = {https://link.aps.org/doi/10.1103/PhysRevLett.117.090402}
}

@article{Yao2017,
  title = {Discrete Time Crystals: Rigidity, Criticality, and Realizations},
  author = {Yao, N. Y. and Potter, A. C. and Potirniche, I.-D. and Vishwanath, A.},
  journal = {Phys. Rev. Lett.},
  volume = {118},
  issue = {3},
  pages = {030401},
  numpages = {6},
  year = {2017},
  month = {Jan},
  publisher = {American Physical Society},
  doi = {10.1103/PhysRevLett.118.030401},
  url = {https://link.aps.org/doi/10.1103/PhysRevLett.118.030401}
}

@article{Zhang2017,
  author  = {Zhang, J. and others},
  title   = {Observation of a Discrete Time Crystal},
  journal = {Nature},
  volume  = {543},
  pages   = {217--220},
  year    = {2017},
  doi     = {10.1038/nature21413}
}

@article{Choi2017,
  author  = {Choi, Soonwon and others},
  title   = {Observation of Discrete Time-Crystalline Order in a Disordered Dipolar Many-Body System},
  journal = {Nature},
  volume  = {543},
  pages   = {221--225},
  year    = {2017},
  doi     = {10.1038/nature21426}
}

@article{Longhi12,
	author = {Stefano Longhi and Giuseppe Della Valle},
	doi = {10.1364/OL.37.002160},
	journal = {Opt. Lett.},
	month = {Jun},
	number = {11},
	pages = {2160--2162},
	publisher = {OSA},
	title = {Anyons in one-dimensional lattices: a photonic realization},
	url = {http://www.osapublishing.org/ol/abstract.cfm?URI=ol-37-11-2160},
	volume = {37},
	year = {2012}}

@article{GreschnerSantos15,
	author = {Greschner, Sebastian and Santos, Luis},
	doi = {10.1103/PhysRevLett.115.053002},
	issue = {5},
	journal = {Phys. Rev. Lett.},
	month = {Jul},
	numpages = {5},
	pages = {053002},
	publisher = {American Physical Society},
	title = {Anyon Hubbard Model in One-Dimensional Optical Lattices},
	url = {https://link.aps.org/doi/10.1103/PhysRevLett.115.053002},
	volume = {115},
	year = {2015}}

@article{Eckardt16,
	author = {Str\"ater, Christoph and Srivastava, Shashi C. L. and Eckardt, Andr\'e},
	doi = {10.1103/PhysRevLett.117.205303},
	issue = {20},
	journal = {Phys. Rev. Lett.},
	month = {Nov},
	numpages = {6},
	pages = {205303},
	publisher = {American Physical Society},
	title = {Floquet Realization and Signatures of One-Dimensional Anyons in an Optical Lattice},
	url = {https://link.aps.org/doi/10.1103/PhysRevLett.117.205303},
	volume = {117},
	year = {2016}}

@article{ZhangGreschner17,
	author = {Zhang, Wanzhou and Greschner, Sebastian and Fan, Ernv and Scott, Tony C. and Zhang, Yunbo},
	doi = {10.1103/PhysRevA.95.053614},
	issue = {5},
	journal = {Phys. Rev. A},
	month = {May},
	numpages = {10},
	pages = {053614},
	publisher = {American Physical Society},
	title = {Ground-state properties of the one-dimensional unconstrained pseudo-anyon Hubbard model},
	url = {https://link.aps.org/doi/10.1103/PhysRevA.95.053614},
	volume = {95},
	year = {2017}}

@article{Greschner18,
	author = {Greschner, Sebastian and Cardarelli, Lorenzo and Santos, Luis},
	doi = {10.1103/PhysRevA.97.053605},
	issue = {5},
	journal = {Phys. Rev. A},
	month = {May},
	numpages = {7},
	pages = {053605},
	publisher = {American Physical Society},
	title = {Probing the exchange statistics of one-dimensional anyon models},
	url = {https://link.aps.org/doi/10.1103/PhysRevA.97.053605},
	volume = {97},
	year = {2018}}

@article{YuanFan17,
	author = {Yuan, Luqi and Xiao, Meng and Xu, Shanshan and Fan, Shanhui},
	doi = {10.1103/PhysRevA.96.043864},
	issue = {4},
	journal = {Phys. Rev. A},
	month = {Oct},
	numpages = {6},
	pages = {043864},
	publisher = {American Physical Society},
	title = {Creating anyons from photons using a nonlinear resonator lattice subject to dynamic modulation},
	url = {https://link.aps.org/doi/10.1103/PhysRevA.96.043864},
	volume = {96},
	year = {2017}}

@article{Batchelor2007,
doi = {10.1088/1742-5468/2007/03/P03007},
url = {https://doi.org/10.1088/1742-5468/2007/03/P03007},
year = {2007},
month = {mar},
publisher = {},
volume = {2007},
number = {03},
pages = {P03007},
author = {Batchelor, M T and Guan, X-W and He, J-S},
title = {The Bethe ansatz for 1D interacting anyons},
journal = {Journal of Statistical Mechanics: Theory and Experiment}
}

@article{BatchelorGuan2008,
doi = {10.1088/1751-8113/41/35/352002},
url = {https://doi.org/10.1088/1751-8113/41/35/352002},
year = {2008},
month = {jul},
publisher = {},
volume = {41},
number = {35},
pages = {352002},
author = {Batchelor, M T and Guan, X-W and Kundu, A},
title = {One-dimensional anyons with competing δ-function and derivative δ-function potentials},
journal = {Journal of Physics A: Mathematical and Theoretical}
}

@article{Slavnov1989,
  author  = {Slavnov, N. A.},
  title   = {Calculation of scalar products of wave functions and form factors in the framework of the algebraic {B}ethe ansatz},
  journal = {Theoretical and Mathematical Physics},
  volume  = {79},
  number  = {2},
  pages   = {502--508},
  year    = {1989},
  doi     = {10.1007/BF01016531}
}

@article{Slavnov1990,
  author       = {N. A. Slavnov},
  title        = {Nonequal-time correlator of currents in the one-dimensional Bose gas},
  journal      = {Theoretical and Mathematical Physics},
  volume       = {82},
  number       = {3},
  pages        = {389--401},
  year         = {1990},
  doi          = {10.1007/BF01029288}
}

@article{Piroli2020,
  author  = {Piroli, Lorenzo and Scopa, Stefano and Calabrese, Pasquale},
  title   = {Determinant formula for the field form factor in the anyonic Lieb--Liniger model},
  journal = {Journal of Physics A: Mathematical and Theoretical},
  volume  = {53},
  number  = {40},
  pages   = {405001},
  year    = {2020},
  doi     = {10.1088/1751-8121/ab94ed}
}

@article{Syrwid20PRR,
  title = {Lack of a genuine time crystal in a chiral soliton model},
  author = {Syrwid, Andrzej and Kosior, Arkadiusz and Sacha, Krzysztof},
  journal = {Phys. Rev. Res.},
  volume = {2},
  issue = {3},
  pages = {032038(R)},
  numpages = {5},
  year = {2020},
  month = {Aug},
  publisher = {American Physical Society},
  doi = {10.1103/PhysRevResearch.2.032038},
  url = {https://link.aps.org/doi/10.1103/PhysRevResearch.2.032038}
}

@incollection{Mack2002,
  author    = {Mack, Holger and Bienert, Marc and Haug, Florian and Straub, Frank S. and Freyberger, Matthias and Schleich, Wolfgang P.},
  title     = {Wave Packet Dynamics and Factorization of Numbers},
  booktitle = {Experimental Quantum Computation and Information},
  editor    = {De Martini, Francesco and Monroe, Christopher},
  series    = {Proceedings of the International School of Physics ``Enrico Fermi''},
  volume    = {148},
  pages     = {369--384},
  publisher = {IOS Press},
  year      = {2002},
  doi       = {10.3254/978-1-61499-004-8-369}
}

@article{Marzoli1998,
  author  = {Marzoli, I. and Saif, F. and Bia{\l}ynicki-Birula, I. and Friesch, O. M. and Kaplan, A. E. and Schleich, W. P.},
  title   = {Quantum Carpets Made Simple},
  journal = {Acta Physica Slovaca},
  volume  = {48},
  number  = {3},
  pages   = {323--333},
  year    = {1998},
  url     = {https://arxiv.org/abs/quant-ph/9806033}
}

@article{Bruno2013Comment,
  title     = {Comment on ``Quantum Time Crystals''},
  author    = {Bruno, Patrick},
  journal   = {Phys. Rev. Lett.},
  volume    = {110},
  number    = {11},
  pages     = {118901},
  year      = {2013},
  publisher = {American Physical Society},
  doi       = {10.1103/PhysRevLett.110.118901}
}

@article{Wilczek2013Reply,
  title     = {Wilczek Reply:},
  author    = {Wilczek, Frank},
  journal   = {Phys. Rev. Lett.},
  volume    = {110},
  number    = {11},
  pages     = {118902},
  year      = {2013},
  publisher = {American Physical Society},
  doi       = {10.1103/PhysRevLett.110.118902}
}

@article{Altuntas2025,
  author        = {Altunta{\c{s}}, E. and Lena, R. G. and
                   Flannigan, S. and Daley, A. J. and Spielman, I. B.},
  title         = {Dynamical structure factor from weak measurements},
  journal       = {Quantum Science and Technology},
  volume        = {10},
  number        = {3},
  pages         = {035045},
  year          = {2025},
  doi           = {10.1088/2058-9565/ade335}
}

@misc{Altuntas2026,
  author        = {Altunta{\c{s}}, Emine and Spielman, Ian B.},
  title         = {Repeated weak measurements: watching quantum
                   correlations evolve},
  year          = {2026},
  eprint        = {2604.17719},
  archivePrefix = {arXiv},
  primaryClass  = {quant-ph},
  doi           = {10.48550/arXiv.2604.17719},
  url           = {https://arxiv.org/abs/2604.17719}
}

@article{VanHove1954,
  author        = {Van Hove, L{\'e}on},
  title         = {Correlations in Space and Time and {Born}
                   Approximation Scattering in Systems of
                   Interacting Particles},
  journal       = {Physical Review},
  volume        = {95},
  number        = {1},
  pages         = {249--262},
  year          = {1954},
  doi           = {10.1103/PhysRev.95.249}
}

@article{Kanamoto2003,
  title = {Stability of the quantized circulation of an attractive Bose-Einstein condensate in a rotating torus},
  author = {Kanamoto, Rina and Saito, Hiroki and Ueda, Masahito},
  journal = {Phys. Rev. A},
  volume = {68},
  issue = {4},
  pages = {043619},
  numpages = {12},
  year = {2003},
  month = {Oct},
  publisher = {American Physical Society},
  doi = {10.1103/PhysRevA.68.043619},
  url = {https://link.aps.org/doi/10.1103/PhysRevA.68.043619}
}

@article{Meinert2015,
  title     = {Probing the Excitations of a Lieb-Liniger Gas from Weak to Strong Coupling},
  author    = {Meinert, F. and Panfil, M. and Mark, M. J. and Lauber, K. and Caux, J.-S. and N{\"a}gerl, H.-C.},
  journal   = {Phys. Rev. Lett.},
  volume    = {115},
  pages     = {085301},
  year      = {2015},
  publisher = {American Physical Society},
  doi       = {10.1103/PhysRevLett.115.085301}
}

@article{Mourigal2013,
  title     = {Fractional spinon excitations in the quantum {H}eisenberg antiferromagnetic chain},
  author    = {Mourigal, Martin and Enderle, Mechthild and Kl{\"o}pperpieper, Axel and Caux, Jean-S{\'e}bastien and Stunault, Anne and R{\o}nnow, Henrik M.},
  journal   = {Nat. Phys.},
  volume    = {9},
  pages     = {435--441},
  year      = {2013},
  publisher = {Nature Publishing Group},
  doi       = {10.1038/nphys2652}
}

@article{Ohberg2019,
  author  = {{\"O}hberg, Patrik and Wright, Ewan M.},
  title   = {Quantum Time Crystals and Interacting Gauge Theories in Atomic {Bose-Einstein} Condensates},
  journal = {Phys. Rev. Lett.},
  volume  = {123}, pages = {250402}, year = {2019},
  doi     = {10.1103/PhysRevLett.123.250402}
}

@article{Syrwid2020Comment,
  author  = {Syrwid, Andrzej and Kosior, Arkadiusz and Sacha, Krzysztof},
  title   = {Comment on ``{Quantum} Time Crystals and Interacting Gauge Theories in Atomic {Bose-Einstein} Condensates''},
  journal = {Phys. Rev. Lett.},
  volume  = {124}, pages = {178901}, year = {2020},
  doi     = {10.1103/PhysRevLett.124.178901}
}

@article{Ohberg2020Reply,
  title = {\"Ohberg and Wright Reply:},
  author = {{\"O}hberg, Patrik and Wright, Ewan M.},
  journal = {Phys. Rev. Lett.},
  volume = {124},
  issue = {17},
  pages = {178902},
  numpages = {2},
  year = {2020},
  month = {Apr},
  publisher = {American Physical Society},
  doi = {10.1103/PhysRevLett.124.178902},
  url = {https://link.aps.org/doi/10.1103/PhysRevLett.124.178902}
}

@article{Syrwid2021,
  author  = {Syrwid, Andrzej and Kosior, Arkadiusz and Sacha, Krzysztof},
  title   = {Can a bright soliton model reveal a genuine time crystal for a finite number of bosons?},
  journal = {EPL},
  volume  = {134}, pages = {66001}, year = {2021},
  doi     = {10.1209/0295-5075/134/66001}
}

@article{Ohberg2024,
  title = {Simulation of time-crystal-like behavior for a few-boson chiral soliton model in a ring},
  author = {{\"O}hberg, Patrik and Wright, Ewan M.},
  journal = {Phys. Rev. A},
  volume = {110},
  issue = {1},
  pages = {013315},
  numpages = {10},
  year = {2024},
  month = {Jul},
  publisher = {American Physical Society},
  doi = {10.1103/PhysRevA.110.013315},
  url = {https://link.aps.org/doi/10.1103/PhysRevA.110.013315}
}

@article{Ohberg2026,
  title = {Quantum time crystals and interacting chiral gauge theories in atomic Bose-Einstein condensates: Role of the Page-Wootters mechanism},
  author = {{\"O}hberg, P. and Wright, E. M.},
  journal = {Phys. Rev. A},
  volume = {113},
  issue = {2},
  pages = {023304},
  numpages = {8},
  year = {2026},
  month = {Feb},
  publisher = {American Physical Society},
  doi = {10.1103/1m4d-kd46},
  url = {https://link.aps.org/doi/10.1103/1m4d-kd46}
}

\newpage
\onecolumngrid
\section{Anyonic density form factor}
The density form factor in the anyonic Lieb–Liniger model can be obtained from the corresponding result for its bosonic counterpart as detailed in the following.
We recall that the anyonic $N$-body wavefunction $\Psi_A$ can be related to a bosonic wavefunction $\Phi_B$ through the mapping \cite{Kundu1999}
\begin{equation}\label{anyon_boson_mapping_wavefunction}
    \Psi_A(x_1,\dots,x_N)=
    \exp\left[-\frac{i\pi\kappa}{2}\sum_{1\leq j<\ell\leq N}\epsilon(x_j-x_\ell)\right]\Phi_B(x_1,\dots,x_N),
\end{equation}
where $\Phi_B$ is symmetric under permutations of its arguments, while the phase factor ensures that $\Psi_A$ satisfies the anyonic exchange relation,
\begin{equation}\label{anyonic_exchange_wave_function}
    \Psi_A(\dots,x_i,x_{i+1},\dots)
    =
    e^{-i\pi\kappa\epsilon(x_i-x_{i+1})}
    \Psi_A(\dots,x_{i+1},x_i,\dots).
\end{equation}
On a ring of length $L$, Eq. (\ref{anyonic_exchange_wave_function}) prevents us from imposing ordinary periodicity in every coordinate \cite{Piroli2020}. In fact, choosing the first coordinate to be periodic consistently gives 
\begin{equation}\label{periodicity_anyonic_wave_function}
\left\{
\begin{array}{l}
\Psi_A(0,x_2,\dots,x_N)
=\Psi_A(L,x_2,\dots,x_N),\\[6pt]
\Psi_A(x_1,0,\dots,x_N)
=e^{-2\pi i\kappa}\Psi_A(x_1,L,\dots,x_N),\\[2pt]
\multicolumn{1}{c}{\vdots}\\[2pt]
\Psi_A(x_1,\dots,x_{N-1},0)
=e^{-2\pi i\kappa(N-1)}
\Psi_A(x_1,\dots,x_{N-1},L).
\end{array}
\right.
\end{equation}
Using the mapping in Eq. (\ref{anyon_boson_mapping_wavefunction}), the boundary conditions in Eqs. (\ref{periodicity_anyonic_wave_function}) imply twisted boundary conditions for the symmetric wavefunction $\Phi_{B}$: 
\begin{equation}\label{boundary_conditions_Phi_B}
  \Phi_{B}(x_{1}, \cdots,0, \cdots, x_{N})=e^{-i\pi\kappa(N-1)}  \Phi_{B}(x_{1}, \cdots,L, \cdots, x_{N}).
\end{equation}
The contact interaction in the Lieb–Liniger Hamiltonian also imposes matching conditions on $\Psi_A$ where particles meet. In particular, for each adjacent pair, integrating the Schrödinger equation across $x_{j+1}=x_j$ for $j=1,\cdots,N-1$ gives 
\begin{equation}\label{jump_anyons_Nbody}
    [(\partial_{x_{j+1}}-\partial_{x_j})\Psi_{A}]|_{x_{j+1}=x_j^{+}}
    -
    [(\partial_{x_{j+1}}-\partial_{x_j})\Psi_{A}]|_{x_{j+1}=x_j^{-}}
    =
    2c\Psi_{A}|_{x_{j+1}=x_j}.
\end{equation}
Using Eqs. (\ref{anyonic_exchange_wave_function}) and  (\ref{anyon_boson_mapping_wavefunction}), these conditions reduce to
\begin{equation}\label{bc_anyons_effective_Nbody}
    [(\partial_{x_{j+1}}-\partial_{x_j})\Phi_{B}]|_{x_{j+1}=x_j^{+}}
    =
    c'\Phi_{B}|_{x_{j+1}=x_j},
    \qquad
    c'=\frac{c}{\cos(\pi\kappa/2)}.
\end{equation}
The eigenvalue problem for $\Psi_{A}$ thus reduces to a bosonic Lieb-Liniger problem for $\Phi_{B}$, with effective coupling $c^{\prime}$ and twisted boundary conditions (\ref{boundary_conditions_Phi_B}).
The wavefunction $\Phi_{B}$ can then be constructed using the coordinate Bethe ansatz. In the ordered domain $0<x_{1}<\cdots<x_{N}<L$, it takes the form $\Phi_{B}(x_{1},\cdots,x_{N})=\sum_{P}A(P)e^{i\sum_{j=1}^{N}\lambda_{P_{j}}x_{j}}$, where $P$ runs over all permutations of the $N$ rapidities $\lambda_{1},\cdots,\lambda_{N}$. The contact conditions (\ref{bc_anyons_effective_Nbody}) determine the relative amplitudes $A(P)$, while enforcing Eq. (\ref{boundary_conditions_Phi_B}) constrains the rapidities through the Bethe equations
\begin{equation}
    e^{i\lambda_{j}L}=e^{i\pi\kappa(N-1)}\prod_{l\neq j}\frac{\lambda_{j}-\lambda_{l}+ic^{\prime}}{\lambda_{j}-\lambda_{l}-ic^{\prime}}.
\end{equation}
The phase factor $e^{i\pi\kappa(N-1)}$ in the Bethe equations reflects the twisted boundary conditions on $\Phi_{B}$.
The twist can however be removed by defining
\begin{equation}\label{periodic_bosonic_wavefunction}
    \widetilde{\Phi}_{B}(x_{1},\cdots,x_{N})
    =
    \exp\left[-\frac{i\pi\kappa(N-1)}{L}
    \sum_{j=1}^{N}x_{j}\right]
    \Phi_{B}(x_{1},\cdots,x_{N}).
\end{equation}
As can be readily checked, $\widetilde{\Phi}_{B}$ is symmetric, periodic in every coordinate, and satisfies the same contact condition
(\ref{bc_anyons_effective_Nbody}) as $\Phi_{B}$.
Moreover, multiplying by the phase factor in Eq. (\ref{periodic_bosonic_wavefunction}) shifts each rapidity to $\tilde{\lambda}_{j}=\lambda_{j}-\pi\kappa(N-1)/L$. These shifted rapidities satisfy the usual periodic Lieb-Liniger Bethe equations with coupling $c^{\prime}$.
We now use the mapping from the anyonic problem to the bosonic problem, given by Eqs. (\ref{anyon_boson_mapping_wavefunction}) and (\ref{periodic_bosonic_wavefunction}), to evaluate the anyonic density form factor.
We label two $N$-particle anyonic eigenstates by the rapidity sets ${\lambda}$ and ${\mu}$ entering the Bethe ansatz for their corresponding bosonic wavefunctions $\Phi_B$. The associated periodic wavefunctions $\widetilde{\Phi}_B$ have shifted rapidity sets ${\widetilde{\lambda}}$ and ${\widetilde{\mu}}$. 
We then have 
\begin{equation}
\begin{aligned}
    \langle\mu|\rho(x)|\lambda\rangle
    &=
    N\int_{0}^{L}dx_{2}\cdots dx_{N}\,
    \Psi_{A}^{*}(x,x_{2},\cdots,x_{N};\mu)
    \Psi_{A}(x,x_{2},\cdots,x_{N};\lambda)\\
    &=
    N\int_{0}^{L}dx_{2}\cdots dx_{N}\,
    \widetilde{\Phi}_{B}^{*}(x,x_{2},\cdots,x_{N};\widetilde{\mu})
    \widetilde{\Phi}_{B}(x,x_{2},\cdots,x_{N};\widetilde{\lambda}),
\end{aligned}
\end{equation}
where the second equality follows from the cancellation of the phases relating $\Psi_{A}$ to $\widetilde{\Phi}_{B}$. The anyonic density form factor therefore equals its periodic bosonic counterpart at coupling $c^{\prime}$, evaluated at $\widetilde{\mu}$ and $\widetilde{\lambda}$.
We can then use the known bosonic expression \cite{Slavnov1989, CalabreseCaux2006}, which depends only on rapidity differences and is therefore unchanged by the common shift.
Written in terms of ${\lambda}$ and ${\mu}$, it gives, at $x=0$,
\begin{equation}\label{anyonic_density_form_factor}
    \langle \mu| \rho(0) |\lambda\rangle:=\Sigma_{\mu,\lambda}
    =
    i^{N}(P_{\mu}-P_{\lambda})
    \prod_{j,k=1}^{N}
    \frac{\lambda_{j}-\lambda_{k}+ic^{\prime}}{\mu_{j}-\lambda_{k}}
    \dt{N-1}{U(\{\mu\},\{\lambda\})}.
\end{equation}
Here, the matrix elements of $U$ are
\begin{equation}
    U(\{\mu\},\{\lambda\})_{j,k}
    =
    \frac{V_{j}^{+}-V_{j}^{-}}{i}\delta_{jk}
    +
    \frac{\prod_{m=1}^{N}(\mu_{m}-\lambda_{j})}
         {\prod_{m\neq j}(\lambda_{m}-\lambda_{j})}
    \left[K(\lambda_{j}-\lambda_{k})-K(\lambda_{N}-\lambda_{k})\right],
\end{equation}
with
\begin{equation}
    V_{j}^{\pm}
    =
    \prod_{m=1}^{N}
    \frac{\mu_{m}-\lambda_{j}\pm ic^{\prime}}
         {\lambda_{m}-\lambda_{j}\pm ic^{\prime}},
    \qquad
    K(z)=\frac{2c^{\prime}}{z^{2}+(c^{\prime})^{2}}.
\end{equation}

\subsection{Specialization to \(N\)-string states}
For our purposes, it is useful to consider Eq. (\ref{anyonic_density_form_factor}) in the attractive regime, with $\bar{c}'=-c'>0$, taking $\ket{\lambda}$ to be an $N$-string state with rapidities
\begin{equation}\label{N_string_lambdas}
\lambda_j=q+\frac{i\bar{c}'}{2}(N+1-2j)+i\delta_j,
    \qquad j=1,\dots,N,
\end{equation}
where $q$ is the string center and $\delta_j$ are exponentially small string deviations. The state $\ket{\mu}$ is instead a generic $N$-particle string state, with rapidities
\begin{equation}
    \mu_{\alpha}^{j,a}
    =\mu_{\alpha}^{j}+\frac{i\bar{c}'}{2}(j+1-2a)
    +i\delta_{\alpha}^{j,a},
    \qquad a=1,\dots,j,
\end{equation}
where $\alpha$ labels different strings of same length $j$ with centers $\mu_{\alpha}^{j}$, and $\delta_{\alpha}^{j,a}$ are the corresponding string deviations. Following the derivation in \cite{CalabreseCaux2006} we obtain, in the ideal string limit,
\begin{equation}\label{Sigma_anyon_specialized_N_string}
|\Sigma_{\mu,\lambda}|
=\frac{N\,\Gamma^{2}(N)\,\big(P_{\mu}-P_{\lambda}\big)^{2}/\bar{c}^{\prime}}{\displaystyle\prod_{j,\alpha,a}
\left[\left(\frac{\mu_{\alpha}^{j}-q}{\bar{c}^{\prime}}\right)^{2}
+\left(\frac{N+j}{2}-a\right)^{2}\right]}\,,
\end{equation}
where $P_{\mu}$ and $P_{\lambda}$ are the momenta of the Bethe states $\ket{\mu}$ and $\ket{\lambda}$, respectively.

\section{Large-N form of the equal-time N-string density-density correlator}
Let us derive the $N$-string contribution to the equal-time density-density correlator at large $N$. From Eq.~(\ref{S_N_x_t}),
\begin{equation}\label{rho_x_t_rho_0_0_sum}
    \langle \rho(x,0) \rho(0,0) \rangle_{N}=\frac{N^{2}}{L^{2}}\sum_{k}\frac{e^{ikx}}{\prod_{a=1}^{N-1}\left[\left(\frac{k}{\cbar^{\prime}Na}\right)^2 +1\right]^{2}},
\end{equation}
with $k=2\pi n/L$, $n\in\mathbb Z$. Applying the Poisson summation formula, $\sum_{n \in \mathbb{Z}} f(n)=\sum_{m \in \mathbb{Z}} \int_{-\infty}^{\infty}dt\,f(t)\,e^{-2\pi i m t}$, to the sum over $n$, and rescaling the integration variable as $u=2\pi t/(\cbar^{\prime}NL)$, we obtain
\begin{equation}\label{rho_poisson}
\langle \rho(x,0) \rho(0,0) \rangle_{N}=\frac{\cbar^{\prime}N^{3}}{2\pi L}\sum_{m \in \mathbb{Z}} \int_{-\infty}^{\infty} du\,\frac{e^{i\cbar^{\prime}N u(x-mL)}}{\prod_{a=1}^{N-1}\left[\left(\frac{u}{a}\right)^2 +1\right]^{2}}.
\end{equation}
The $m=0$ term gives the profile for $|x|\le L/2$, while the terms with $m\neq0$ are its periodic images, centered at $x=mL$. Since the profile decays on the scale $1/(\cbar^{\prime}N)$, for $L\gg 1/(\cbar^{\prime}N)$ the images are negligible in this interval, and we retain only $m=0$.
Using $\sinh(\pi u)/(\pi u)=\prod_{a=1}^{\infty}\left[(u/a)^2+1\right]$, the product in the  integrand of Eq.~(\ref{rho_poisson}) can be written as
\begin{equation}
   \frac{1}{\prod_{a=1}^{N-1}\left[\left(\frac{u}{a}\right)^2 +1\right]^{2}}=\left(\frac{\pi u}{\sinh \pi u}\right)^{2} \prod_{a=N}^{\infty}\left[\left(\frac{u}{a}\right)^2+1\right]^{2}.
\end{equation}
For large $N$ at fixed $u$, expanding the logarithm of the remaining product and using $\sum_{a=N}^{\infty}1/a^{2}=1/N+1/(2N^{2})+O(N^{-3})$, we find
\begin{equation}
2\sum_{a=N}^{\infty}\log\left[\left(\frac{u}{a}\right)^{2}+1\right]=\frac{2u^{2}}{N}+\frac{u^{2}}{N^{2}}+O(N^{-3}),
\end{equation}
and hence
\begin{equation}
\prod_{a=N}^{\infty}\left[\left(\frac{u}{a}\right)^{2}+1\right]^{2}=1+\frac{2u^{2}}{N}+\frac{u^{2}+2u^{4}}{N^{2}}+O(N^{-3}).
\end{equation}
This expansion is valid for $u^2\ll N$. Since the factor $(\pi u/\sinh\pi u)^2$ exponentially suppresses larger values of $u$, the expansion can nevertheless be integrated over the entire real line. We therefore obtain
\begin{equation}
\label{eqS41}
\langle \rho(x,0) \rho(0,0) \rangle_{N}=\frac{\cbar^{\prime}N^{3}}{2\pi L}\left[G(z)-\frac{2}{N}G^{(2)}(z)+\frac{2G^{(4)}(z)-G^{(2)}(z)}{N^{2}}+O(N^{-3})\right],\qquad z=\cbar^{\prime}N x,
\end{equation}
where $G^{(n)}=d^{n}G/dz^{n}$ and
\begin{equation}
    G(z)=\int_{-\infty}^{\infty}du\,e^{iuz}\left(\frac{\pi u}{\sinh\pi u}\right)^{2}=\pi\,\frac{\frac{z}{2}\cosh\frac{z}{2}-\sinh\frac{z}{2}}{\sinh^{3}\frac{z}{2}}.
\end{equation}
The leading term of (\ref{eqS41}) gives the expression for $S_{N}(x,0)$ in Eq. (\ref{N_string_bump_profile})  
quoted in the main text.

\section{Multi-string contributions to the density-density correlation function}
Here, we extend the calculation of the
ground-state density-density correlation function
$\langle \rho(x,t)\rho(0,0)\rangle$ in the attractive anyonic Lieb--Liniger model beyond the $N$-string contribution given in Eq.~(\ref{S_N_x_t}) of the main text, by including intermediate states in which the $N$-string is broken into shorter strings. Using the Lehmann representation, we have
\begin{equation}
\label{density_density_correlator_lehmann}
    \langle \rho(x,t)\rho(0,0)\rangle
    =
    \sum_{\mu}
    e^{i(P_{\mu}-P_{\text{GS}})x-i(E_{\mu}-E_{\text{GS}})t}
    \frac{
        |\langle \text{GS}|\rho(0,0)|\mu\rangle|^2
    }{
        \norm{\text{GS}}^2 \norm{\mu}^2
    },
\end{equation}
where the sum is performed over all intermediate $N$-particle  string states $\ket{\mu}$. 
The density form factor $\langle \text{GS}| \rho(0,0)|\mu \rangle$ is given in Eq. (\ref{Sigma_anyon_specialized_N_string}). The momentum and energy of the state $\ket{\mu}$ are \cite{CalabreseCaux2006}
\begin{equation}
    P_{\mu}=\sum_{j,\alpha} j\mu_{\alpha}^{j} ,\quad E_{\mu}=\sum_{j,\alpha}\left[j(\mu_{\alpha}^{j})^{2}-\frac{\cbar^{\prime 2}}{12}j(j^{2}-1)\right],
\end{equation}
where $\mu_{\alpha}^{j}$ denotes the center of the $\alpha$-th string of length $j$, as explained in the main text.
Moreover, the norm of such states is given by the Gaudin-Korepin formula \cite{Gaudin2014,Korepin1982,CalabreseCaux2006, Piroli2020}
\begin{equation}
    \norm{\mu}^{2}=\cbar^{\prime N_{s}} \prod_{j} j^{N_{j}} \prod_{(k,\beta)>(j,\alpha)} F_{jk}(\mu_{\alpha}^{j}-\mu_{\beta}^{k}) \det G^{(r)},
\end{equation}
where $N_{j}$ is the number of strings of length $j$ in the state $\ket{\mu}$, $N_{s}=\sum_{j}N_{j}$ is the total number of strings, and the second product runs over all distinct pairs of strings. The function $F$ is defined as 
\begin{equation}
    F_{jk}(\lambda)=\frac{\lambda^{2}+(\frac{\cbar^{\prime}}{2}(j+k))^{2}}{\lambda^{2}+(\frac{\cbar^{\prime}}{2}(j-k))^{2}},
\end{equation}
and $G^{(r)}$ is the reduced Gaudin matrix
\begin{equation}
    G^{(r)}_{(j,\alpha),(k,\beta)}=\frac{\partial}{\partial \mu_{\beta}^{k}}\left[j\mu_{\alpha}^{j}L-\sum_{(l,\gamma)}\Phi_{jl}(\mu_{\alpha}^{j}-\mu_{\gamma}^{l})\right],
\end{equation}
with
\begin{equation}
    \Phi_{jk}(\lambda):=(1-\delta_{jk})\phi_{|j-k|}(\lambda)+2\phi_{|j-k|+2}(\lambda)+\dots+ 2\phi_{j+k-2}(\lambda)+\phi_{j+k}(\lambda), \quad \phi_{m}(\lambda):=2\arctan{\frac{2\lambda}{m \bar{c}^{\prime}}}.
\end{equation}
Finally, the allowed string centers $\mu_{\alpha}^{j}$ are determined by the Bethe equations
\begin{equation}
    j\mu_{\alpha}^{j}L-\sum_{(k,\beta)\neq (j,\alpha)} \Phi_{jk}(\mu_{\alpha}^{j}-\mu_{\beta}^{k})-\pi \kappa (N-1)j = 2\pi\left(n+\frac{N_{j}-1}{2}\right), \quad n\in \mathbb{Z}.
\end{equation}

\end{document}